\documentclass[%
 reprint,
 amsmath,amssymb,
 aps,
]{revtex4-2}

\usepackage{graphicx}
\usepackage{dcolumn}
\usepackage{bm}

\begin{document}

\preprint{APS/123-QED}

\title{On the Crystallization Kinetics in Natural Rubber}
\author{Rabia Laghmach}
\email{laghmach@iastate.edu}
 \affiliation{Department of Chemistry, Iowa State University, Ames IA, 50011, USA}
 \affiliation{Institut Lumière Matière, UMR5306 Université Lyon 1-CNRS\\ Université de Lyon 69622 Villeurbanne cedex, France}
\author{Nicolas Candau}
\affiliation{
 Centre Català del Plàstic (CCP), Universitat Politècnica de Catalunya (UPC)\\
 Barcelona Tech (EEBE-UPC), Av. D'Eduard Maristany, 16, 08019, Spain 
}%
\author{Laurent Chazeau}
\affiliation{
INSA-Lyon, Université Lyon I, MATEIS CNRS UMR5510, 69621, France 
}%
\author{Thierry Biben}%
 \email{thierry.biben@univ-lyon1.fr}
\affiliation{%
Institut Lumière Matière, UMR5306 Université Lyon 1-CNRS\\
Université de Lyon 69622 Villeurbanne cedex, France
}


\date{\today}

\begin{abstract}
In this article, we introduce a framework to investigate the growth of nano-crystallites in a polymer matrix numerically. This framework combines the Flory theory of entropic elasticity with phase-field approaches commonly used to model crystal growth. We investigate in particular the growth kinetics of a crystallite in the presence of topological constraints such as entanglements or cross-links, and show that depending on the coupling between the topological constraints and the growth kinetics various structures can be observed: branched structures looking like spherulites or stable nano-crystallites as observed in strain-induced crystallization.
\end{abstract}

\maketitle
Although a large number of studies have been devoted to Natural Rubber (NR), its fascinating properties are still a matter of debate. NR is an elastomer made of natural polyisoprene (cis-1,4-polyisoprene) that is cross-linked during the vulcanization process. The polymer network produced during such treatment has the ability to resist very large deformations (it can be reversibly elongated by a factor higher than 7). An important property for industrial applications is its ability to crystallize under strain, providing Natural Rubber a self-reinforcement behavior. Despite many years of experimental and theoretical studies, a clear understanding of the time-dependent crystallization mechanism and its consequence on the mechanical properties is still lacking. However, the possibilities offered by modern X-ray facilities now allow an investigation of the crystallites growth at the nanometric scale. It is our scope to discuss the theoretical counterpart in this article.

Crystallization under strain is usually understood in the framework suggested by Flory~\cite{Flory47}, where elasticity originates from entropic considerations. This entropic nature of the elasticity gives elastomers a behavior under strain that differs significantly from more conventional (atomic) solids, where enthalpy is the usual source of elasticity~\cite{Treloar75, Wall51, Edwards_1988}. Crystallization is favored by an applied strain precisely because the formation of a crystal relaxes entropy and, consequently, stress. However, many features related to crystallization kinetics are not understood: a simple explanation of why the crystallites have a nanometer size when cyclic strain is applied~\cite{murakami2002} or can form branched structures at a very long time in cold crystallization experiments is still lacking~\cite{Andrews1962, Andrews1964, Andrews1971, TOKI2000, Trabelsi2003, Poompradub2005, TOSAKA2012, CANDAU2012}. In this paper, we shall explore the kinetics of crystallization in NR using a local (microscopic) thermodynamic approach. We shall use Flory’s ideas and test different models accounting for the topological constraints imposed both by the presence of the cross-links and the physical entanglements. Since our scope is to test different scenarios (kinetic limitation of the growth due to the topological constraints or elastic limitation), we shall use simplified models that can be easily refined to better description of the SIC phenomena. Although simple, these models are able to give predictions on the growth kinetics of the crystallites, on their shapes and sizes, with a good quantitative agreement with experimental findings when a comparison is possible.

To introduce the used framework, we shall focus on a single crystal nucleus and discuss its behavior under strain and various temperature conditions. The amorphous to crystal phase transition is described using a local expression for the Flory entropy of crystallization. Following Flory, the free enthalpy is vanishing in the crystal phase, taken as the reference, and has two contribu- tions in the amorphous phase: an enthalpic contribution that fixes the melting temperature to Tm in the absence of applied strain, and the entropic elastic contribution when local deformations are present (such as the deformation induced by the crystal formation or by an applied strain). Introducing $\theta({\bf{r}},t)$ the local crystal fraction at point $r$ and time $t$ ( $\theta = 0$ in the amorphous phase and $1$ in the crystal phase), the Flory theory writes~\cite{Flory47}:
\begin{equation}
\begin{array}{rl}
f_{\text{bulk}}(\theta=1)= & 0\text{ in the crystal phase (reference phase)}\\
f_{\text{bulk}}(\theta=0)= & \nu\left(nh_{f}\frac{T_{m}-T}{T_{m}}+k_{B}T\:\text{tr}\mathbf{E}\right)\text{ (amorphous)}
\end{array}\label{eq:bulk}
\end{equation}
where $\nu$ is the active network chain density, $n$ is the average number of segments between two cross-links, $h_{f}$ is the melting enthalpy per segment, $k_{B}$ is the Boltzmann constant, $T$ is temperature and $\mathbf{E}$ is the local strain tensor in the large deformation regime (\emph{i.e. }it contains all the nonlinear contributions~\cite{Landau86}). This expression corresponds to Flory's calculation for the elastic response of the amorphous phase considered as incompressible, for the uniaxial elongation by a factor $\lambda$ considered by Flory we indeed recover $\text{tr}\mathbf{E}=\lambda^{2}/2+1/\lambda-3/2$.
One also can show that in the small deformation regime $\text{tr}\mathbf{E}\to\text{tr}\left(\mathbf{\epsilon}^{2}\right)$ where $\mathbf{\epsilon}$ is the small deformation strain tensor, as expected for an incompressible system. 
\begin{figure}
\centering
\includegraphics[width=1.0\columnwidth]{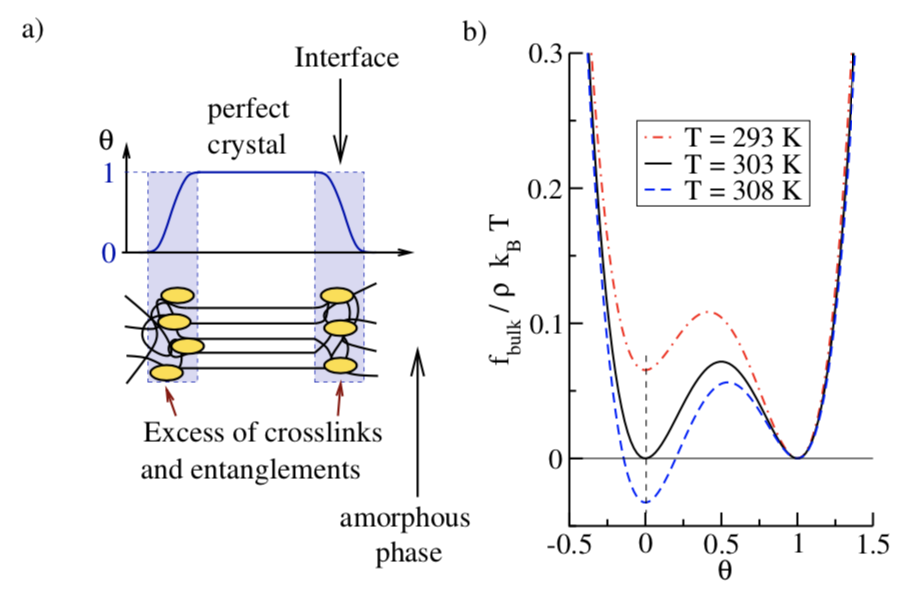}
\caption{a) Elementary ingredients of the model. b) Bulk free enthalpy functional (2) for $T_{m} = 303K$}
\label{fig:Geometry}
\end{figure}
In the interfacial region, the crystal fraction may vary
rapidly between $0$ and $1$ as illustrated in figure 1a), and it is necessary to know the behavior of $f_{\text{bulk}}(\theta)$ between these two values. Since the two bulk phases correspond to minima of the bulk free enthalpy,
$f_{\text{bulk}}(\theta)$ has a maximum between $\theta=0$ and $\theta=1$. In the spirit of phase-field models, we assume a simple double well form for $f_{\text{bulk}}(\theta)$, whose true shape is actually unknown:
\begin{equation}
\begin{array}{rl}
f_{\text{bulk}}(\theta)=\Gamma\frac{\theta^{2}}{4}(1-\theta)^{2} & +g(\theta)\nu\left(nh_{f}\frac{T_{m}-T}{T_{m}}+k_{B}T\:\text{tr}\mathbf{E}\right)\\
\text{with:} & g(\theta)=1-\theta^{2}(3-2\theta)
\end{array}\label{eq:fbulk}
\end{equation}
One can check that the bulk expressions~(\ref{eq:bulk}) are verified. In this expression, $\Gamma$ is an energy density scale that controls the energy barrier between the crystal and the amorphous phase at coexistence. This parameter is related to the surface tension as we shall see later. The particular cubic shape of $g(\theta)$ is chosen to ensure that its
minimum values are always at $\theta=0$ and $\theta=1$, the equilibrium bulk phases, whatever the value of temperature $T$~\cite{Kassner2001} (see Figure 1b). For the inhomogeneous system we wish to describe, a gradient expansion of the free enthalpy functional is required to get a consistent description of the interfaces~\cite{Rowlinson82}. We use here the standard square gradient expansion, which is a simple way to introduce the surface tension of the nucleus, an important component of the theory of nucleation:
\[
F[\theta,\mathbf{u}]= \int\int\int d{\bf r}\left[ f_{\text{bulk}}(\theta,\mathbf{u}) + \frac{w^{2}}{2}\left(\nabla\theta\right)^{2} \right]
\]
with the prescription, minimization of the functional $F[\theta,\mathbf{u}]$ at coexistence between the crystal and the amorphous phase produces smooth interfaces of thickness $w$ and surface tension $\gamma=w\int_{0}^{1}\sqrt{2\Gamma f_{\text{bulk}}(\theta)} d\theta$~\cite{Rowlinson82}. An analytical expression can be obtained using (\ref{eq:fbulk}): $\gamma=w\Gamma/6\sqrt{2}$. Consequently, we can adjust both $w$ and $\Gamma$ to match the experimental interface thickness and the surface tension with our model. 

With such ingredients, and a dynamic equation for the local order parameter $\theta({\bf{r}},t)$, it is possible to solve the dynamics of the nucleus growth, without assumption on its shape. The used dynamic equation is an Allen--Cahn kinetic equation~\cite{ALLEN19791085, Hohenberg1977}:
\begin{equation}
\frac{\partial\theta(\mathbf{r},t)}{\partial t}=  -\alpha_{\theta}\:\frac{\delta F}{\delta\theta(\mathbf{r},t)}
\label{eq:kinet}
\end{equation}
that relaxes the order parameter field $\theta$ towards the minimum value of the free enthalpy. $\alpha_{\theta}$ is a kinetic coefficient that fixes the relaxation time $\tau_{\theta}$ of the order parameter field $\theta$ to its equilibrium value. $\tau_{\theta}$, the only time scale in the model, will be used as a unit scale. We shall discuss its value later. Equation (\ref{eq:kinet}) must be solved with the condition of local mechanical equilibrium at any time: 
\begin{equation}
\nabla.\mathbf{\sigma}=0
\label{eq:mecha}
\end{equation}
where $\sigma(\mathbf{r})$ is the local stress tensor. Doing so, inertia is neglected at the scale of the nucleus, a reasonable assumption for nanometer scale crystals. To numerically achieve this condition, it is replaced with a second kinetic equation, introducing the local deformation field $\mathbf{u}(\mathbf{r})$:
\begin{equation}
\frac{\partial\mathbf{u}(\mathbf{r})}{\partial t}=  -\alpha_{\mathbf{u}}\:\frac{\delta F}{\delta\mathbf{u}(\mathbf{r})} \equiv \alpha_{\mathbf{u}} \nabla.\mathbf{\sigma}
\label{eq:kinet2}
\end{equation}
where $\alpha_{\mathbf{u}}$ is again a kinetic constant that defines the relaxation time $\tau_{\mathbf{u}}$ of the deformation field. Taking a value of $\tau_{\mathbf{u}}$ much smaller than $\tau_{\theta}$ (at least a factor of 10) ensures a rapid relaxation of the stress so that equation (\ref{eq:mecha}) is satisfied at any time step.

Let us summarize the physical parameters already present at this stage: the phase transition itself is governed by the Flory theory that introduces several control parameters such as the monomer density $\rho=1.47\:10^{4}\text{mol/m}^{3}$, and $n$ the number of segments between two cross-links. The active network chain density is defined as $\nu=\rho/n$. The value of $n$ depends on the cross-link density, and we used $n=95$ to compare with available experimental data~\cite{CANDAU2012}. The melting temperature is a controversial parameter $T_{m}=250$K, the value used by Flory~\cite{Flory47}, corresponds to the most rapid growth rate of the crystallites without applied strain~\cite{Wood1946}, however, the melting temperature of an infinite crystal as been reported as large as $308$K~\cite{DALAL1983}. As a compromise, $T_{m}=308$K is used in this work. Following Flory, the melting enthalpy $h_{f}$ per segment is taken as $h_{f}=600 k_{b}$~\cite{Flory47}. The parameter $\Gamma$ is fixed such that the interfacial tension between the amorphous phase and the crystal is $2.10^{-2}\text{J/m}{}^{2}$~\cite{DALAL1983}. The interface thickness $w$ is fixed to $w=1\:\text{nm}$, a compromise between the size of a segment ($0.3\:\text{nm}$) and the size of a crystallite (few nanometers). 

\begin{figure}
\includegraphics[width=1.0\columnwidth]{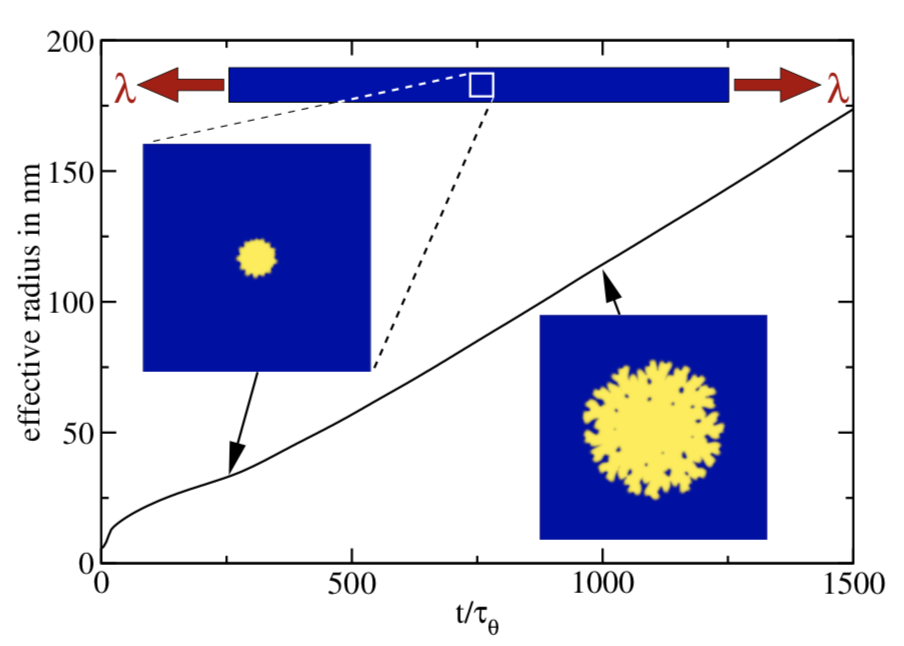}
\caption{\label{fig:KinetRad}Effective radius of the nucleus as a function of time for the kinetic model ($m=5$, $\lambda=1$, i.e no applied strain). The images show the geometry of the crystallite at different time steps, the amorphous phase is blue while yellow is the crystal. The morphological instability allows a faster growth of the nucleus.}
\end{figure}

Once the values of the parameters have been fixed, and an initial configuration is chosen for $\theta(\mathbf{r})$ and $\mathbf{u}(\mathbf{r})$, the coupled equations (\ref{eq:kinet}) and (\ref{eq:kinet2}) can be solved numerically. At each time step, a simple Euler scheme is used with a finite difference method on a cubic or square grid to evaluate the spatial derivatives. The initial state in all the simulations presented below consists in an homogeneous portion of the amorphous phase stretched by a factor $\lambda$ in the $x$-direction. The simulation box is a small square of dimension $400\:\text{nm} \times 400\:\text{nm}$ at the center of the sample, schematized as a white square on the top part of figure~\ref{fig:KinetRad}. This box is filled with a square grid of lattice spacing $1\:\text{nm}$ in each direction. A vertex $(x,y)$ of the grid corresponds to the positions \emph{after} deformation. The problem is solved in 2 dimensions to follow the large scale dynamics of the nucleus in a reasonable time; it somehow mimics a thin film. One can easily obtain the value of the homogeneous deformation field that satisfy the incompressibility condition, $u_{x}= \frac{\lambda-1}{\lambda}x$ and $ u_{y}= (1-\lambda)y$. At time $t=0$, a small circular crystallite of radius $r_{n}$ between $5\:\text{nm}$ and $7.5\:\text{nm}$ is generated in the center of the grid. Its dynamics is obtained by solving iteratively the equations of the model. The crystallite can grow or melt depending on the conditions. If its size is below the critical nucleation radius, it melts. To generate the nucleus we used the formula:
\[
\theta(r)=\frac{1}{2}\left\{ 1-\tanh\left(\frac{r-r_{n}}{2\sqrt{2}w}\right)\right\} 
\]
where $r=\sqrt{x^{2}+y^{2}}$. The boundary conditions are $\theta=0$ at the edge of the rectangular box, $u_{x}=\frac{\lambda-1}{\lambda}L_{x}/2$ on the right side, $u_{x}=-\frac{\lambda-1}{\lambda}L_{x}/2$ on the left side, $u_{y}=(1-\lambda)L_{y}/2$ on the top side and $u_{y}=-(1-\lambda)L_{y}/2$ on the bottom side.  

Without additional constraints trivial results are obviously obtained, such as an unlimited growth of the nucleus below the melting temperature or above a critical elongation ratio. Experimentally, crystallite growth is known to be limited, likely because the topological constraints created by the crosslinks and the entanglements are expelled from the crystal and pushed away in the amorphous phase. However, the precise mechanism and its consequences are not known at the moment. These topological constraints can be introduced in the phase-field framework as a local density $\rho_{\text{topo}}(\mathbf{r})$. As a first step, we consider that these topological constraints can be described by a conservation equation:
\[
\frac{\partial\rho_{\text{topo}}}{\partial t}+\nabla.\mathbf{J}=0
\]
where $\mathbf{J}$ is the local flux. Since by definition the crystal phase is well ordered, we assume here that no topological defect is present inside the crystal domain (defects are possible in this model, but they will simply not correspond to $\theta=1$). The model thus assumes that the topological defects are expelled from the crystallite during the growth. The flux $\mathbf{J}$ is thus localized in the interfacial region where the crystal phase is produced. We can thus set $\mathbf{J}=\rho_{\text{topo}}\mathbf{v}$ where $\mathbf{v}$ is the interface velocity. This velocity corresponds to the velocity of the frame that follows the interface, the interface should thus be steady in such frame and $\mathbf{v}$ should satisfy:
\[
\frac{\partial\theta}{\partial t}+\mathbf{v}.\nabla\theta=0 \rightarrow \mathbf{v}=-\frac{\partial\theta}{\partial t}\frac{\nabla\theta}{\left\Vert \nabla\theta\right\Vert ^{2}}
\]
With this prescription, topological constraints are gathered in the vicinity of the interface. Indeed, an excess of entanglements or crosslinks in the interfacial region is expected to penalize the formation of a pure crystalline phase, but the precise mechanism and its consequences are not known at the moment. In this work, we propose two possible mechanisms and follow the effects on the structure: a kinetic effect, where the topological constraints slow down the growth by increasing locally $\tau_{\theta}$, or an elastic effect due to the deformation of the crosslink and/or the entanglement network, where the local excess of elastic energy should unfavor the crystal phase. 
\begin{figure}
\includegraphics[width=1.0\columnwidth]{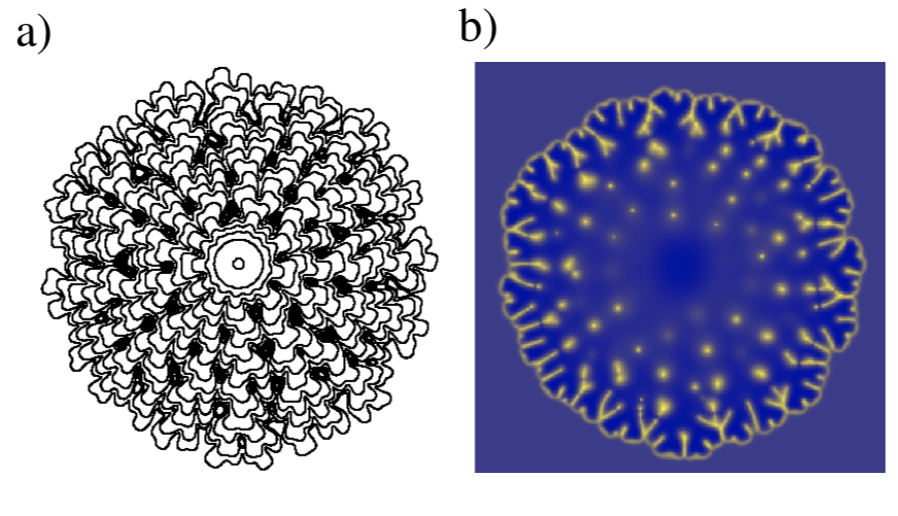}
\caption{\label{fig:Morphology}a) Morphology of the crystallite at different time steps (every 100$\tau_{\theta}$) for the kinetic model. b) Relative density of the topological constraints at time $1400\tau_{\theta}$, inside the crystal no entanglement or cross-link remain while around the interface an enrichment as large as $300\%$ is observed. Far away, the distribution of topological constraints is not modified.}
\end{figure}

\begin{figure}
\includegraphics[width=1.0\columnwidth]{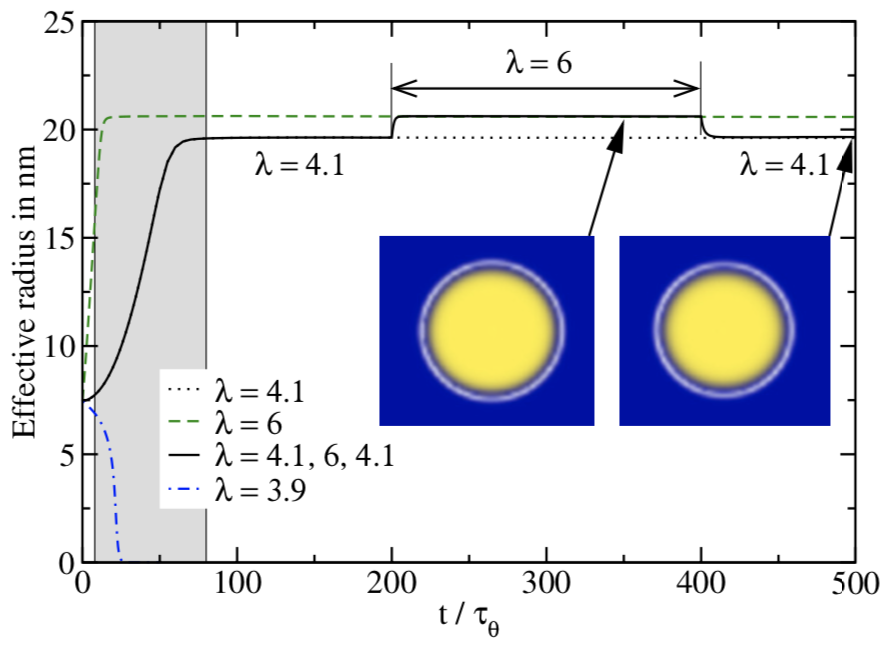}
\caption{\label{fig:ElasticModel}Effective radius as a function of time for a crystallite at room temperature ($300K$) as obtained with the Elastic model. When $\lambda<4$ the nucleus melts while above $4$ the nucleus grows to reach an equilibrium shape.}
\end{figure}
The "Kinetic Model" assumes that the phase transition is delayed by the accumulation of topological constraints in the vicinity of the interface. To implement this idea, $\tau_{\theta}$ is simply rescaled by a factor $1+(\rho_{\text{topo}}/\rho_{\text{topo}}^{\infty})^{m}$, where $\rho_{\text{topo}}^{\infty}$ is the density of topological constraint far away from the crystallite. The exponent $m$ is varied between 1 and 5 aiming at stopping the growth. However, even if this description is able to predict a slow down of the growth quite efficiently initially, the system finds a solution by producing a branched structure. The resulting crystalline phase has a spherulitic shape induced by the rejection of the topological constraints on the lateral sides so that the growth can continue in the radial direction. This is illustrated in Fig.~\ref{fig:KinetRad} and Fig.~\ref{fig:Morphology} where the evolution of the effective radius (the square root of the surface divided by $\pi$) is plotted (Fig.~\ref{fig:KinetRad}). After a fast transient, the growth slows down until the branching instability begins (the top image of Fig.\ref{fig:KinetRad}). Next, the growth speeds up again until it reaches a nearly constant velocity and the crystallite invades the whole system. Fig.\ref{fig:Morphology}-a) shows the morphology of the crystallite at different time steps; we can follow the emergence of the branches and their subsequent splitting at the tip, a mechanism that differs from the usual dendrite instability where the secondary branches are formed on the side of the main branch~\cite{Langer1980}. Fig.\ref{fig:Morphology}-b) gives a view of the topological constraints that are rejected between the branches and remain trapped there. 

Considering now the "Elastic model", a deformation field is associated to the entanglement and/or the cross-link network. We compute the displacement field $\mathbf{u}_{\text{topo}}$ of these topological constraints thanks to the following transport equation: 
\[ 
\frac{\partial\mathbf{u}_{\text{topo}}}{\partial t}+\mathbf{v}.\nabla\mathbf{u}_{\text{topo}}=\mathbf{v}
\]
where $\mathbf{v}$ is the growth velocity defined previously. This equation simply expresses that the topological constraints are transported with the velocity $\mathbf{v}$. An additional elastic energy is associated to this displacement field, and for the sake of simplicity we use the small deformation theory to evaluate its effect on the growth (displacements remain at the nanometer scale). We add this elastic energy to the free enthalpy density (\ref{eq:fbulk}): 
\[
f_{\text{bulk}} \rightarrow -\frac{g(\theta)}{2}\mathbf{\sigma}_{\text{topo}}.\mathbf{\epsilon}_{\text{topo}}
\]
where $\mathbf{\epsilon}_{\text{topo}}$ is the strain tensor associated to the displacement field $\mathbf{u}_{\text{topo}}$ in the small deformation regime, and $\mathbf{\sigma}_{\text{topo}}$ the corresponding stress tensor. Please note that the topological constraints being rejected in the amorphous phase, their additional elastic contribution is restricted to this phase where $g(\theta)= 1$. In the previous model, an enrichment of $\rho_{\text{topo}}$ in the interfacial region as large as $300\%$ was observed. This may lead to a Young modulus for the additional elastic contribution between one and three times the Young modulus of the amorphous elastomer. However, we expect the enrichment to be more limited here, due to the saturation of the crystallite size. Therefore, the value of the additional Young modulus for the enriched zone was chosen equal to $500$kPa, slightly above the Young modulus of the amorphous elastomer. With this value the growth is stopped very rapidly. The crystallite shape remains almost circular but the effective radius of the crystallite depends on the traction that is applied (as can be seen in figure~\ref{fig:ElasticModel}). The effective radius of a $w=7.5\:\text{nm}$ nucleus as a function of time is plotted in figure~\ref{fig:ElasticModel}: we can see that below a critical elongation $\lambda_c = 4$, the nucleus simply melts and only elongations larger than $4.1$ can give rise to a stable crystallite. With the chosen parameters, the computed $\lambda_c$ is very close to the experimental one: $\lambda_c \approxeq 4$~\cite{CANDAU2012}. The shapes are shown for $\lambda=4.1$ (right image) and $\lambda=6$ (left image). After a sudden increase of the elongation from $\lambda=4.1$ up to $\lambda=6$ at time $t =200\tau_{\theta}$ (the black curve), the crystallite is able to resume its growth until it reaches its new equilibrium; more interestingly, this transformation is reversible: reducing $\lambda$ from $6$ to $4.1$ (at time $t =400\tau_{\theta}$, we can see that the crystallite melts to go back to its original shape. A quantitative investigation of the kinetics shows that a crystallites equilibrates its shape with time between $8$ and $80\tau_{\theta}$ (the grey area in figure~\ref{fig:ElasticModel}). This gives an estimation of $\tau_{\theta}$ to be at most of the order of $2.5\:\text{ms}$, in agreement with the experimental growth times of the crystallites (between $20\:\text{ms}$ and $200\:\text{ms}$)~\cite{CANDAU2012}. 

To conclude, we have shown that the two limiting mechanisms investigated for the strain-induced crystallization phenomenon, namely kinetic or elastic limitation of the growth, are leading to very different structures. Growth instabilities producing spherulite-like shapes are observed when entanglements and/or crosslinks are delaying the growth (kinetic limitation) while the accumulation of elastic constraints in the neighborhood of the cristallites are on the contrary stopping it. In reality we expect both mechanisms to play a role, but the interplay between the two still needs to be quantified. Many other features as well as improved prescriptions for the kinetics and the thermodynamics can be tested with the framework we propose. For instance the coupled structural and mechanical response of NR to a cyclic strain is presently investigated.


\bibliography{apssamp}

\end{document}